\documentclass[12pt]{article}
\usepackage{color,epsfig}
\usepackage{graphicx}
\definecolor{violet}{rgb}{0.4,0,0.2}
\definecolor{vert}{rgb}{0,0.6,0.0}
\definecolor{navy}{rgb}{0.0,0.0,0.6}
\definecolor{orange}{rgb}{0.6,0.4,0.0}
\definecolor{bleu}{rgb}{0.3,0.0,0.8}
\definecolor{spin}{rgb}{0.3,0.6,0.0}

\def\vv{{\color{violet}v}} 
 \def\phhi{{\color{violet}\varphi}}
\def\calK{{\color{violet}{\cal K}}}\def\sS{{\color{vert}S}}

\def\nn{{\color{vert}n}}
\def\kk{{\color{vert}k}}

\def\mm{{\color{red}m}}
\def\microm{{\color{red}m^{_\oplus}}}
\def\ee{{\color{red}e}}\def\jj{{\color{red}j}}
\def\mmu{{\color{red}\mu}}\def\hba{{\color{red}\hbar}}
\def\sigm{{\color{red}\sigma}}
\def\Elec{{\color{black}\rm E}}
\def\UU{{\color{red}U}}  \def\VV{{\color{red}V}}
\def\calH{{\color{red}{\cal H}}}

\def\pp{{\color{red}p}}
\def\calE{{\color{red}{\cal E}}}
\def\varepsi{{\color{red}{\varepsilon}}}
\def\alp{{\color{violet}\alpha}}

\def\calR{{\color{black}{\bf T }}}

\def\bfK{{\color{vert}{\bf K }}}

\def\calV{{\color{bleu}{\cal V }}}
\def\gl{{\color{vert}{\ell }}}
\def\Bl{{\color{black}{\ell }}}
\def\le{{\color{bleu}{e }}}
\def\lle{{\color{black}{ l }}}
\def\ita{{\color{bleu} {\it a} }}
\def\itb{{\color{bleu} {\it b} }}

\def\bb{\large\color{black}  $ }  \def\fb{ $  }
\def\be{\large\begin{equation}\color{black} }
\def\fe{\end{equation}}
\def\rf{\color{black} (\ref }
\def\fr{)\,\color{navy} }

\def\rmn{ {\rm n}} \def\rmp{{\color{orange}\rm p}}

\def\spose#1{\hbox to 0pt{#1\hss}}\def\lta{\mathrel{\spose{\lower 3pt\hbox
{$\mathchar"218$}}\raise 2.0pt\hbox{$\mathchar"13C$}}}  \def\gta{\mathrel
{\spose{\lower 3pt\hbox{$\mathchar"218$}}\raise 2.0pt\hbox{$\mathchar"13E$}}}

\begin{document}

\title{\color{navy}\bf Entrainment coefficient and effective mass for
conduction neutrons in neutron star crust : simple microscopic models}

\author { {\bf
Brandon Carter$^{a}$,
Nicolas Chamel$^{a}$,
Pawel Haensel$^{a,b}$ }
\\ \hskip 1 cm\\   \\
$$ $^{a}$ Observatoire de Paris, 92195 Meudon, France \\
(Brandon.Carter@obspm.fr, Nicolas.Chamel@obspm.fr)$$
\\ $$ $^{b}$ N. Copernicus Astronomical Center, Warsaw, Poland \\
 (haensel@camk.edu.pl)$$ }

\maketitle

{\catcode `\@=11 \global\let\AddToReset=\@addtoreset}
\AddToReset{equation}{section}
\renewcommand{\theequation}{\thesection.\arabic{equation}}

\color{navy}

\bigskip
{\bf Abstract.}    In the inner crust of a neutron star, at densities
above the ``drip'' threshold, unbound ``conduction'' neutrons can move 
freely past through the ionic lattice formed by the nuclei. The relative 
current density {\bb\nn^i= \nn \bar \vv^i \fb} of such conduction neutrons will be related
to the corresponding mean particle momentum {\bb\pp_i\fb} by a
proportionality relation of the form {\bb\nn^i= {\calK}\pp^i\fb} in terms
of a physically well defined mobility coefficient {\bb{\calK}\fb} whose
value in this context has not been calculated before. Using methods from
ordinary solid state and nuclear physics, a simple quantum mechanical
treatment based on the independent particle approximation, is used here to formulate
{\bb{\calK}\fb} as the phase space integral of the relevant group velocity
over the neutron Fermi surface. The result can be described as an
``entrainment'' that changes the ordinary neutron mass {\bb\mm\fb} to a
macroscopic effective mass per neutron that will be given -- subject to adoption of a
convention specifying the precise number density {\bb\nn\fb} of the neutrons
that are considered to be ``free'' -- by {\bb\mm_\star=\nn/ {\calK}\, .\fb}
The numerical evaluation of the mobility coefficient is carried out for nuclear configurations of the
``lasagna'' and ``spaghetti'' type that may be relevant at the base of the
crust. Extrapolation to the middle layers of the inner crust leads to the
unexpected prediction that {\bb\mm_\star\fb} will become very large compared
 with {\bb\mm\, .\fb}

\vskip  1 cm

\section{Introduction}

The main purpose of this article is to show how a mean field treatment of neutron star
crust matter can be used  to address the
previously unsolved problem of evaluating a quantity, namely the relevant
neutron mobility coefficient {\bb \calK\fb}, that is essential for the
astrophysical applications that will be described in a separate
article~\cite{CCHII}. In terms of the relevant number density {\bb \nn\fb}
of effectively unbound neutrons, this coefficient determines a corresponding
effective mass {\bb \mm_\star=\nn/\calK\fb} that characterises their average
motion on a macroscopic scale (meaning one that is large compared with
the spacing between nuclei) and that we therefore refer to as the macro mass,
to distinguish it from the microscopic effective mass,
{\bb\microm\fb} say, characterising the dynamics of the neutrons on
subnuclear scales. Whereas {\bb\microm\fb} is well known to be typically
rather smaller than the ordinary neutron mass {\bb\mm\fb}, we reach the
previously unexpected conclusion that there is likely to be a strong
``entrainment'' effect whereby the macro mass {\bb \mm_\star\fb} will
typically become large, and in some layers extremely large, compared
with {\bb\mm\fb}.

A secondary purpose of this article is to draw the attention of nuclear
theorists to the potentialities of the almost entirely unexplored branch
of theoretical astrophysical nuclear physics that needs to be developed for
this and many other purposes. The only relevant work of which we are aware
so far is that of Oyamatsu and Yamada ~\cite{Oyamatsu94}, who appear to be
the only ones to have taken proper account of the neutron scattering by the
nuclei inside the inner crust by the use of appropriate
 Bloch type periodicity conditions of the kind commonly employed
for the treatment of electrons in ordinary terrestrial solid state physics. Their
treatment however was restricted to a simple one dimensional model.

Under the conditions of ordinary terrestrial solid state physics, and even
at the much higher densities characterising the matter in a white dwarf star,
long range electric forces keep the nuclei so far apart that, in so far as
the much stronger but short range nuclear interactions are concerned, each
individual nucleus can be treated separately as if it were isolated. Until
quite recently~\cite{Oyamatsu94}, such a separate treatment of
individual nuclei (considered as if isolated each in its own cell - with
 Wigner Seitz type boundary conditions) has been used in
nearly all quantum mechanical calculations on neutron star crust matter since the pioneer work of Negele
and Vautherin~\cite{Vautherin73}. That kind of approximation is fully
justifiable in the outer crust , where the densities are not too much greater
than those found in a white dwarf. However such a treatment can no longer be
considered entirely satisfactory in the inner crust, meaning the part with
density above the ``neutron drip'' threshold at about 10$^{11}$ g/cm$^3$, where
there are unconfined neutrons that travel between neighbouring nuclei, which
thereby cease to be effectively isolated from one another.

While desirable for accuracy throughout the inner crust, a proper collective
rather than individual treatment of the nuclei becomes not just desirable but
absolutely essential for treating the problem with which Oyamatsu and Yamada
~\cite{Oyamatsu94}~\cite{Oyamatsu93} were concerned, namely that of the
nuclear matter inside neutron star crust.
Such a treatment is also essential for treating the problem with which the
present work is concerned, namely that of stationary but non static
configurations in which a neutron current flows relative to the lattice
formed by the nuclei, something that obviously can not be discussed in the
usual approach that treats the nuclei as if they were isolated in individual
(e.g. Wigner Seitz type) boxes.

The flow of neutrons is treated here as a perturbation of a zero temperature
ground state characterised just by the location of the relevant Fermi surface in
momentum space. We thereby obtain provisional rough estimates of the
relevant mobility coefficient which suggest that (unlike what occurs in
the fluid core and on a microscopic scale) the macro mass {\bb\mm_\star\fb}
that effectively characterises the neutron motion on a macroscopic scale
can become very large compared with the ordinary neutron mass {\bb\mm\fb},
particularly in the middle part of the conducting layer, for which three dimensional
numerical results will be presented in a follow up article~\cite{Chamel04}.
The present article deals more specifically with simplified rod and plate type models
that are relevant near the crust core interface, where the mass enhancement will be
less extreme.

Bulgac, Magierski and Heenen have recently pointed out
the importance of shell effects induced by unbound neutrons in neutron star crust by
evaluating the Casimir energy for neutron matter in the presence of inhomogeneities from a semiclassical approach
and more recently by performing a Skyrme Hartree-Fock calculation
with ordinary periodic boundary conditions (see \cite{BulMagHeen02, MagHeen02} and references therein). However this kind of
boundary conditions does not properly account for Bragg scattering of dripped neutrons and is only a particular case
of the more general Bloch type boundary conditions.

In the absence of any previous quantum mechanical calculation whereby nuclei on a crystal lattice
are treated collectively (apart from the 1D calculation previously mentionned
 \cite{Oyamatsu94}),
even at the simplest level of approximation,
we shall adopt the simple model suggested by Oyamatsu and Yamada, supplemented with Bloch type boundary
conditions, in order to estimate the effective neutron mass {\bb\mm_\star\, .\fb} This model treats
the neutrons as independent fermions subject to an effective
potential. We wish to draw the attention of nuclear theorists to the problem of
including  Bloch boundary conditions in more sophisticated approximation schemes
as a challenge for future work.

In the mean time, experience with the analogous problem of electron transport
in ordinary solid state physics suggests that the results obtained from the
Oyamatsu Yamada type treatment used here should not be too bad as a first
approximation. Further encouragement comes from our own recent attempt to take
up the challenge of allowing for coupling by an appropriate adaptation of the
standard BCS pairing theory on which the prediction of neutron superfluidity
(in the relevant low to moderate temperature range) is based: the
upshot~\cite{CCHIII} is that (although it is essential for the inhibition
of resistivity) as far as the ``entrainment'' phenomenon is concerned the
effect of the ensuing pairing ``gap'' will not be very large nor very difficult to
calculate.

\section{Microscopic description of conduction neutrons in the inner crust}
\label{sect:micro.qphysics}

\subsection{Single-particle Schr\"odinger equation}

The basic principle of the conductivity model we wish to adapt from ordinary
solid state theory to the context of a neutron star crust is that the
``conduction band'', and perhaps also some of the highest ``confined''
levels, can be analysed within the independent particle approximation, in
terms of energy eigenstates for a single particle described by a Bloch type
wave function {\bb \phhi \fb}, satisfying the Bloch periodic boundary
conditions as discussed below. In the rest frame of the crust, with respect
to which the system will be assumed to be in stationary equilibrium, the
single particle wave function will be taken to be governed by a Hamiltonian
operator {\bb \calH\fb} that is given by
{\be \calH =  -\hba ^2\nabla_i \,\frac{\gamma^{ij}}{2\microm}\,
\nabla_j   +\VV \, ,\label{0a}\fe}
where {\bb\gamma^{ij}\fb} is just the space metric, while  {\bb \VV \fb} is
the single particle potential, and {\bb \microm\fb} is the relevant
local effective mass parameter.

The potential {\bb \VV \fb} and the effective mass {\bb \microm\fb} will be periodic in
the case of a regular crystalline solid lattice, though not for a fluid
configuration such as will be relevant at higher temperatures. The value of
the effective potential {\bb \VV \fb} (and the associated deviation of the
effective mass {\bb \microm\fb} from  {\bb\mm\fb}) is supposed to
allow not just for the attraction of the confined protons in the nuclei but
also for the mean effect of the other fermions, which are electrons in the
familiar solid state applications, but neutrons in the context under
consideration here.

\subsection{Boundary conditions}
\label{sect:boundarycond.}

The preceding considerations apply even to disordered (glass like or liquid)
configurations, but in order to proceede we shall now restrict our attention to
cases for which the nuclei are assumed to be fixed and locally distributed
in a regular crystal lattice configuration, in which an elementary cell consists of a parallelopiped of
volume {\bb {\calV}_{\rm cell}\fb} say spanned by a triad of basis vectors
{\bb {\bf \le}_\ita \fb} labelled by an index with values {\bb \ita=\fb}
1, 2, 3, so that the system is invariant with respect to translations
generated by vectors of the form {\be\calR=\Bl^\ita{\bf \le}_\ita
\label{trans}\fe} for integer values of the coefficients {\bb \Bl^\ita\fb} (in the
following summation over repeated indices is assumed). This so called adiabatic
or Born-Oppenheimer approximation is justified by the large difference between the
neutron and nucleus masses,
since typically each nucleus contains several hundred nucleons \cite{DH01}.

 It follows directly from the well-known Floquet-Bloch theorem \cite{Ashcroft81}
  that the single particle  wave function has to satisfy the following boundary conditions
 {\be \phhi_{\bf \kk }\{ {\bf r }+{\bf \calR } \}  = {\rm e}^{{\rm i}\,
 {\bf \kk }\cdot {\bf \calR }}\phhi_{\bf \kk }\{ {\bf r } \} .
 \label{Blochphase}\fe} where {\bb \bf \kk \fb} is the Bloch momentum covector.

It will therefore suffice to solve the Schr\"odinger equation just
inside a single elementary cell. Instead of using a primitive cell of
parallelopiped, it is convenient for many purposes to work instead with the
 Wigner-Seitz (W-S) cell defined as the set of points that are closer to a given
lattice node than to any other. Such a cell exhibits the full symmetry of the
lattice. Its shape is determined by the crystal structure, a polyhedron in
general, for instance a cube in a simple cubic lattice. This exactly defined
 W-S cell should not be confused with the
widely employed eponymous ``W-S approximation'' \cite{WignerSeitz33} which consists
in replacing this cell by a sphere (or more generally any convenient cell that simplifies
the analysis).

The Bloch momentum {\bb \bf \kk \fb} takes values inside the first Brillouin
zone (B-Z), which is the W-S cell of the reciprocal lattice whose nodes
are located at {\bb \bfK = \gl_\ita {\bf \lle}^{\ita}\fb} for integer
values of the coefficients {\bb \gl_\ita \fb} where the dual basis vectors
{\bb {\bf \lle}^{\ita}\fb} are defined by the following scalar products
 {\be {\bf \lle^{\ita}\cdot \le_{\itb} }= 2\pi \delta^\ita_\itb\, ,
\label{dualbase} \fe}
the {\bb 2\pi\fb} normalization factor being introduced for
convenience.

It must be emphasized that the single particle wave function will have to
satisfy the relation (\ref{Blochphase}) between two opposite faces of the cell.
This means in particular that the Schr\"odinger equation has to be solved for
\emph{each} wave vector {\bb \bf \kk \fb} inside the first B-Z. The
ordinary periodic boundary conditions on the cell, which have been recently applied
by Magierski and Heenen \cite{MagHeen02}, are thus only a restricted
subset of solutions, namely those with {\bb {\bf \kk} =0.\fb}

For each momentum {\bb \bf  \kk \fb}, there exists only a discrete set of energy
eigenvalues satisfying the Bloch boundary conditions. The single particle
energy spectrum is therefore a collection of sheets in momentum space, each
sheet usually being referred as a band (labelled by the principal quantum
number).

\section{Microscopic dynamics in mean field of lattice}
\label{micr}
\subsection{From microscopic to macroscopic observables}
\medskip

The linearisation involved in the neglect of direct two body or many body
interactions in such a mean field treatment excludes allowance for higher
order effects such as the pairing responsible for a superfluid energy gap,
but as discussed in a separate article \cite{CCHIII}, this limitation should
not matter too much for the evaluation of the
basic equation of state of the fluid, since the main effect of superfluidity
is not to modify the equation of motion but just to restrain the class of
the admissible solutions (by allowing only those that are irrotational).

The main limitation on the use of such a linearisation is that it makes
sense only for configurations that do not differ too much from the static
reference configuration on which the estimation
of the effective potential energy function {\bb  \VV \fb} is based.

\subsection{Fermi surface in ground state configuration}

The zero temperature configuration is obtained by minimising an energy density
{\bb \UU\fb} subject to the constraint of a given value of the neutron density
{\bb \nn_\rmn \, , \fb} defined by {\be \nn_\rmn =\int {{\rm d}^3 \kk\over (2\pi)^3} \label{2a} \, . \fe}
The energy density is expressible in terms of the single particle
energies (in this work, we shall use braces instead
of ordinary brackets for functional dependence, in order to avoid possibly confusion with
 simple multiplication) {\be \calH \phhi_{\bf \kk }={\calE}\{ {\bf \kk } \}\phhi_{\bf \kk }\, ,
\label{00}\fe}
by an expression of the form
{\be \UU= \UU\{0\} + \int {\calE}\{ {\bf \kk } \} {{\rm d}^3 \kk\over (2\pi)^3} \, ,\label{1}\fe}
where, as a standard postulate, the contribution {\bb \UU\{0\}\fb} is ignored
in the minimisation procedure. This means that all single particle states of energy up to
but not beyond some particular Fermi level, {\bb \mmu\fb} say, are occupied. This Fermi energy
specifies the ``Fermi surface'', namely the locus, {\bb  \sS_{_{\rm F}}\fb} say,
in momentum space, where
{\be {\calE}\{{\bf \kk} \}= \mmu\, .\label{7b}\fe}

It must be emphasised that the Fermi surface will in general consist of
\emph{disconnected} pieces unlike the homogenous case in which the Fermi surface
is simply a sphere.  From the property that {\bb \calE\{ {\bf \kk}\} = \calE\{ -{\bf \kk} \}, \fb}
the ground state is thus completely symmetric in the sense that its total momentum density {\bb  \int {\rm d}^3 \kk\,
\kk^i \fb} vanishes. This is a generalized version of Uns\"old theorem
\cite{Cowan} according to which the all electron wave function of a closed
shell atom is spherically symmetric.

\subsection{Minimal conducting configurations}

Our main concern in the present analysis is with current carrying
configurations, as characterised by some given value of the physically
well defined number current density with components {\bb \nn^i\fb} given
by {\be \nn^i= \int \vv^i {{\rm d}^3 \kk\over (2\pi)^3}\label{2b} \fe}
in which the transport velocity {\bb \vv^i\fb} is given by the formula
{\be \vv^i = \frac{1}{\hba} \frac{\partial \calE}{\partial k_i}\label{6} \, . \fe}

 The kind of current carrying configurations that will
presumably be relevant in the low temperature limit will be those for which
the energy density {\bb \UU\fb} is minimised for a given value of the neutron
density {\bb \nn_\rmn\fb} subject to the further constraint that the number
current density also has the prescribed value {\bb \nn^i.\fb} It is evident
that this constrained minimisation condition requires that the occupied
states should consist just of those subject to an inequality of the form
{\be {\calE}\leq \mmu +\pp_i\, \vv^i\, ,\label{8}\fe}
for some fixed covector whose constant components {\bb \pp_i\fb} have been
introduced as Lagrange multipliers.
The phase space volume specified in this way will have a boundary
characterised by the condition
{\be {\calE}=\mmu+\pp_i\, \vv^i\, ,\label{8b}\fe}
which specifies a modified Fermi surface, {\bb \sS\fb} say. The constraint
of having a finite current breaks the symmetry and will thus necessarily
lead to a new ground state with a non vanishing total momentum density
vector {\bb  \int {\rm d}^3 \kk\, \kk^i \neq 0\fb}.

\subsection{The uniform displacement of the Fermi surface}

\begin{figure}
\centering
\epsfig{figure=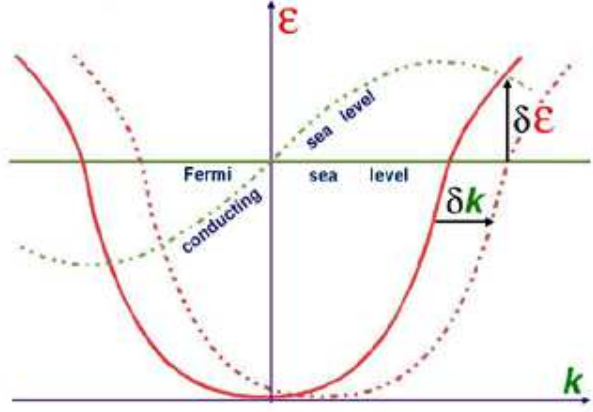, width=8 cm}
\caption{Sketch of energy against wave number, showing uniform displacement
attributable to current.
\label{fig1}}
\end{figure}

To justify the use of a Schr\"odinger type Hamiltonian {\bb \calH\fb}
involving an effective potential of the same form as in the zero current
case, we need to assume that (as will
presumably be the case in the relevant context of pulsar glitches) the
current density {\bb \nn^i\fb} is small enough to be treated by linear perturbation theory.
This means that the Lagrange multiplier {\bb \pp_i\fb} should itself be
considered just as a first order perturbation, having zero value
{\bb \pp_i=0\fb} in the static unperturbed configuration, and {\bb \pp_i
=\delta\pp_i\fb} in the perturbed current carrying configuration. For a
given (unchanged) value of the chemical potential {\bb \mmu\fb}, the
difference between the new value \rf{8b}\fr and the old value \rf{7b}\fr
of the Fermi energy level {\bb \calE\fb} will be given to first order by
{\bb \delta\calE=\delta(\pp_i\,\vv^i)\fb}{\bb  =\vv^i\, \delta\pp_i\fb}
and thus in terms of the perturbed value of {\bb \pp_i\fb} simply by
{\be\delta\calE=\pp_i\,\vv^i\, .\label{12}\fe}
This change in energy can be interpreted as being attributable to a
phase space displacement {\bb \delta \kk_i.\fb} Since this change will be given,
according to \rf{6}\fr, by
{\be \delta {\calE}=\hba \vv^i\delta \kk_i\, ,\label{10}\fe}
it can be seen that the simplest possibility is to take the displacement to
have the {\it uniform} value given just by
{\be \hba\delta\kk_i=\pp_i\, ,\label{14} \fe}
as illustrated on figure \ref{fig1}.

\subsection{Relation between current and momentum}

The relation \rf{14}\fr means that (in the infinitesimal limit) the
Lagrange multipliers {\bb \pp_i\fb} can be physically interpreted simply as
components of a {\it uniform} pseudo momentum displacement of the occupied
phase space region, and of its ``Fermi surface'' boundary.

The effect of this displacement on the Fermi
surface element {\bb {\rm d}\sS_{_{\rm F}}{^i}\fb} as given -- in terms of the
corresponding surface measure element, {\bb {\rm d}\sS_{_{\rm F}}\fb} --
by {\be{\rm d}\sS_{_{\rm F}}{^i}=(\vv^i/\vv){\rm d}\sS_{_{\rm F}}
\, ,\label{11}\fe} will be to sweep out an infinitesimal phase
space volume element given by
{\be\hba {\rm d}^3\kk=\pp_i\, {\rm d}\sS_{_{\rm F}}^{\,i}\,  .\label{15}\fe}
It follows that, for the integral over the occupied region of any phase
space function {\bb  f\, ,\fb} the difference between the value for the
displaced (conducting) configuration and the value for the non conducting
reference configuration will be given (to linear order) by the formula
{\be\hba\, \delta\!\int \! f\, {\rm d}^3 \kk=\pp_i\oint_{_{\rm F}}\! f\,
{\rm d}\sS_{_{\rm F}}^{\,i}\, ,\label{16}\fe}
in which it is to be understood that the integral on the
right is taken over the entire Fermi surface.

We have already pointed out that the energy function {\bb \calE\fb} will
be symmetric with respect to the origin, hence it follows that the
Fermi surface element {\bb {\rm d}\sS_{_{\rm F}}^{\,i}\fb}  will be antisymmetric about the
origin so that its unweighted integral
{\bb  \oint_{_{\rm F}} {\rm d}\sS_{_{\rm F}}^{\,i}\fb} will self cancel
to give zero. We can thus see from \rf{16}\fr that for any phase space
function  {\bb  f\fb} that is symmetric about the origin the corresponding
integral will be unaffected by the displacement, i.e. we shall get
{\bb  \delta\int  f\, {\rm d}^3 \kk=0\fb}. This contrasts with the case of
an antisymmetric function, for which it is the unperturbed integral that
will vanish, i.e. we shall have {\bb \int  f\, {\rm d}^3 \kk=0\fb} in the
static reference configuration, so that the corresponding value for the
conducting configuration will be given by
{\bb \int  f\, {\rm d}^3 \kk=\delta\int  f\, {\rm d}^3 \kk\fb}.

The antisymmetric
case is illustrated by the application
with which this work is principally concerned, namely the current density
{\bb  \nn^i\fb} given by \rf{2b}\fr. Since we have {\bb \nn^i=0\fb} in the
reference configuration characterised by {\bb \pp_i=0\fb}, we shall have
{\bb \nn^i=\delta \nn^i\fb} in the conducting configuration characterised
by {\bb  \pp_i=\delta\pp_i\fb}. Thus by substituting {\bb \vv^i\fb} for
{\bb  f\fb} in \rf{16}\fr we see that in the linearised limit the current
will be related to the pseudo momentum displacement {\bb \pp_i\fb} by a
relation of the form
{\be \nn^i={\calK}^{ij}\pp_j\, ,\label{18}\fe}
in which the symmetric tensor {\bb \calK^{ij}\fb} is given as an integral over the
Fermi surface by
{\be {\calK}^{ij}={1\over (2\pi)^3\hba }\oint_{_{\rm F}}\! {\vv^i \vv^j
\over \vv}\, dS_{_{\rm F}} \, .\label{21}\fe}

Before continuing, it is to be remarked that this tensor {\bb \calK^{ij}\fb} is
interpretable as being proportional to the zero temperature limit of
the electric conductivity tensor {\bb \sigm^{ij},\fb} defined in the usual way by
{\bb \jj^i=\sigm^{ij}\Elec_j\fb} where {\bb  \Elec_j\fb}
is the relevant electric field and {\bb \jj^i\fb} the electric current density,
by a relation of the form {\bb \sigm^{ij}=\tau \ee^2{\calK}^{ij}, \fb}
where {\bb \tau\fb} is the
relevant relaxation time and {\bb \ee \fb} the electric charge per particle.

In a entirely disordered (liquid or glass like) state the form of the
energy function {\bb \calE\fb} and the consequent location in phase space of
the Fermi surface will be difficult to evaluate theoretically, but there
will be the partially compensating simplification that the result will
automatically be isotropic. In the mathematically simpler case of a
cubic crystalline lattice that is expected~\cite{Ruderman68}
to occur in a neutron star crust, this tensor  {\bb \calK^{ij}\fb} will
have the isotropic form
{\be {\calK}^{ij}={\calK}\gamma^{ij}\, ,\label{25}\fe}
in which the scalar coefficient will evidently be given by
{\be {\calK}={_1\over ^3}\gamma_{ij}{\calK}^{ij}\, .\label{26}\fe}
However that may be, it can be seen
from \rf{21}\fr that the scalar coefficient {\bb \calK\fb} will be given by
a Fermi surface integral of the simple form
{\be {\calK}={1\over 3(2\pi)^3\hba}\oint_{_{\rm F}}\!\vv\, {\rm d}
\sS_{_{\rm F}}\, .\label{27}\fe}

In terms of this integral, the relation between current and momentum
will be given by
{\be \nn^i={\calK}\gamma^{ij}\pp_j\, .\label{27b}\fe}

\subsection{The (cut-off dependent) concept of the effective mean mass}
\label{cutoff}

In order to relate the formula \rf{27}\fr to expressions used elsewhere
in the literature, it is to be noted that we can introduce a mean
velocity vector {\bb \bar v^i\fb} and a conduction neutron density {\bb \nn \fb} such that
the current density can be expressed by
{\be \nn^i=\nn\bar v^i\, .\label{28a}\fe}

Subject to the isotropy condition \rf{25}\fr, the (pseudo) momentum
covector {\bb \pp_i\fb} will be expressible in terms of the corresponding
covector {\bb \bar v_i=\gamma_{ij}\bar v^j\fb} in the simple form
{\be \pp_i = \mm_\star\bar v_i\, ,\label{28}\fe}
in which the effective mean particle mass {\bb \mm_\star\fb} --  which we
refer to as the ``macro mass'' in order to distinguish it from the
locally  effective ```micro mass'' {\bb \microm\fb} --
is defined in terms of the integral \rf{27}\fr by the relation
{\be {1\over \mm_\star}={{\calK}\over \nn} \, .\label{29}\fe}
It is however to be remarked that whereas the specification of quantities
such as {\bb \pp_i\fb} and {\bb \calK\fb} is physically unambiguous, the
specification of the effective mass {\bb \mm_\star\fb}, like that of the mean
transport velocity {\bb \bar v^i\fb}, depends on how many ``conduction
states'' are counted in the definition of the number density {\bb  \nn\, ,\fb}
and how many are left aside as dynamically inert ``confined'' states.

In the application to a neutron star crust, the situation is somewhat
simpler than is usual in ordinary solid state physics because the relevant
single particle Hamiltonian \rf{0a}\fr will usually involve an effective potential
function {\bb \VV\fb} that tends rapidly~\cite{Haensel00} towards an almost
exactly uniform value outside the
ionic nuclei to which all the protons and some of the neutrons are effectively
 confined.  It should be noticed that the single particle wave function of
bound states will be vanishingly small at the W-S cell boundary and therefore those
states will not be sensitive to the Bloch phase shift. The resulting
single particle energies will thus be nearly independent of the Bloch momentum
which means that the associated group velocities will be very small. The single
particle energy spectrum can therefore be decomposed into a subset consisting of
such confined states, and a remainder that will realistically be describable
as ``conduction'' states. The separation between those two different subsets
is not entirely sharp (borderline states are commonly refered to as valence states
in ordinary solid state physics) and one has to rely on a more or less arbitrary
convention.

In most layers of a neutron star crust, the most convenient possibility is usually
to take the uniform value of {\bb \VV\fb} outside nuclei,
which can be taken as the energy origin, to specify a corresponding
energy range {\bb {\calE} > 0\fb} characterising ``conduction'' states. This 
specification may however be ambiguous in the bottom layers of the inner
crust where nuclei are very close to each other. In order to deal with such cases,
we shall adopt the convention, which agrees with the previous one when nuclei are
very far apart, that conduction states are defined as states whose
energy is larger than the maximum value of the potential.

It is to be remarked that in the uniform limit whereby the periodic
potential {\bb \VV \fb} and the local effective mass {\bb \microm \fb} are simply
constants, the resulting Fermi surface is a sphere and the evaluation of the mobility scalar
is straightforward. In this case, the associated macro mass is just equal to the micro mass
{\bb \mm_\star = \microm \fb} provided all neutron states are counted as conduction states.

\section{Quantitative estimates of the mobility tensor}
\subsection{Bottom layer of the inner crust}
\begin{figure}
\centering
\epsfig{figure=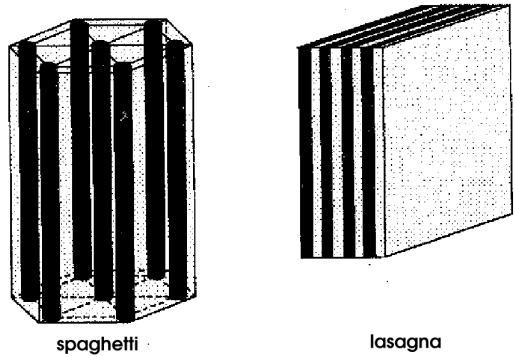, height=5 cm}
\caption{Nuclear configurations for hexagonal spaghetti (rod) and lasagna
(slab) lattices.
Black - nuclear matter; grey - neutron gas. From \cite{Oyamatsu93}, with
kind permission of the author.
\label{fig3}}
\end{figure}

We shall focus for simplicity on the bottom of the inner crust near saturation density
of the order of $10^{14}$ g/cm$^3$ where the transition to
homogeneous fluid neutron matter takes place, to evaluate the mobility scalar and the
effective mass. In this region the crystal lattice
is not expected to substantially alter the transport
properties such as the mobility tensor. The aim of the following sections is
not to provide astrophysically important information but rather to
give some insight that may be valuable in the more general cases that will be
dealt with in future work. Near the base of the crust, the nuclei are so
strongly deformed by their neighbours that they may adopt non spherical
shapes \cite{PR95}, sometimes idealised by 1D or 2D configurations such as slabs and
rods respectively, which greatly simplifies the analysis. These ``exotic''
crust phases are  illustrated on figure \ref{fig3} taken from the work of
Oyamatsu \cite{Oyamatsu93}.

In views of the exploratory character of the present work, we shall use a simple model for the single particle equation. In particular,
we shall drop the condition of strict self-consistency,
and shall use a mean field model based on the work of Oyamatsu \cite{Oyamatsu93, Oyamatsu94}. He calculated the structure of the ground state
of the inner neutron star
crust using a phenomenological energy-density functional,
fitted on the first hand to the smoothed nuclear masses and charged radii of laboratory nuclei on the $\beta$ stability line
and on the other hand adjusted on the equation of state of pure neutron matter from the variational calculations of
Friedman and Pandharipande \cite{Friedman81} using two body as well as three body nucleon-nucleon interactions.
He further investigated with Yamada the importance of shell effects with a single particle Schr\"odinger equation in
the W-S approximation \cite{Oyamatsu94} (the reader is refered to those two papers
for the details).

Unlike the calculations carried out by these authors,
we shall apply the 1D or 2D Bloch type boundary conditions and we shall not
make any approximations about the shape of the W-S
cells which partition the crystal. The 3D configurations require specific numerical techniques
and will be discussed in a separate
work \cite{Chamel04}. For consistency, the lattice parameters (whose values are found in \cite{Oyamatsu93})
will be defined so that the volume of the exact W-S cell is equal
to the volume of the approximate cell. Within a W-S cell, we approximate the single 
particle potential {\bb  \VV\fb} by the potential {\bb  \UU\fb} of Oyamatsu and Yamada \cite{Oyamatsu94}. A
zeroth order approximation {\bb  \UU_0\fb} of the potential is obtained by differentiation of the potential part of an energy density functional
{\bb v\{\nn_\rmn, \nn_\rmp\} \fb} so that
{\be \UU_0 = \frac{\partial v}{\partial \nn_\rmn }. \fe} Specifically we shall
use the parameter set of model I of this paper. Then, the effect of the finite range of the nucleon-nucleon interaction
is taken into account by using the folding of the potential {\bb \UU_0 \fb}
with a gaussian smearing function of width {\bb \kappa. \fb}
This potential is supplemented with a spin-orbit
coupling term, which is parametrized as a functional of the density gradients. The gaussian width and the parameters
of the spin-orbit potential are adjusted so as to reproduce the correct sequence of single particle energy levels
of $^{208}$Pb.
The potential {\bb \UU \fb} thus obtained varies only near the nuclear
surface. Therefore, close to the W-S
cell boundary {\bb \UU\fb} is constant. This enables us to express the periodic
potential {\bb \VV\fb} acting on a neutron moving in an infinite lattice of
nuclei as
{\be
\VV \{ {\bf r} \}=\sum_{\calR } \UU\{ {\bf r }-{\calR } \}~,
\label{U.lattice}
\fe}
where the sum goes over all nuclei occupying  the lattice sites, i.e.
{\bb {\bf \calR } = \Bl^\ita {\bf \le}_\ita \fb}, where {\bb \Bl^\ita\fb} are
integers. The potential {\bb \VV\{ {\bf r } \}\fb} thus possesses all
the symmetries of the crystal lattice. We neglect small changes in
{\bb \UU\fb} due to passing from a single W-S cell to an infinite lattice. The energy origin
is taken as the value of the potential {\bb \VV \fb} outside nuclei.
Following Oyamatsu and Yamada \cite{Oyamatsu94} we do not include momentum dependent
terms in the single particle potential. Consequently the microscopic effective neutron mass in this model is
just equal to the bare neutron mass {\bb \microm = \mm\,.\fb}
We have ignored the spin-orbit coupling since it is found to be
about one order of magnitude smaller than the central
potential \cite{Oyamatsu94}.

The single-particle Schr\"odinger equation is solved here by the Rayleigh-Ritz variational
approach whereby the expectation value of the Hamiltonian is minimized
subject to a normalization condition. The single particle wave function is
expanded into plane waves defined by
{\be \phhi_{\bf \kk } \{ {\bf r } \} = \frac{1}{\sqrt{\calV_{\rm cell}}}
\sum_{\bfK } \tilde{\phhi}_{\bf \kk }\{ {\bfK } \}\, {\rm e}^{ {\rm i}\,
({\bf \kk} + \bfK )\cdot{\bf r}}\, ,\label{Blochpacket}\fe} with the
 normalization
{\be \sum_{\bfK } |\tilde{\phhi}_{\bf \kk }({\bfK })|^2 =1\, .\fe}

This expansion into plane waves is actually exact, it merely makes more
apparent the periodicity. The approximation lies in the fact that the
summation needs to be truncated for practical calculations at some cut off
energy {\bb \calE_{\rm cut off}\fb} (which thus makes this method adequate
only for slowly varying potentials as in the present case), i.e. so as to
include all reciprocal lattice vectors satisfying
{\be \frac{\hba^2 ({\bf \kk +\bfK })^2}{2 \mm} < \calE_{\rm cut off} \, .\fe}

The group velocity is found according to the Hellmann-Feynman theorem
\cite{Feynman39} to be equal to
{\be {\bf \vv} = \frac{\hba {\bf \kk }}{\mm} +\sum_\bfK \frac{\hba \bfK}
{\mm} |\tilde{\phhi}_{\bf \kk }({\bfK})|^2\, .\label{groupvel}\fe}
 This last equation shows that deviations from the homogeneous case arise
from the spreading of the Bloch wave packet (\ref{Blochpacket}).

For each total neutron density {\bb \nn_{\rmn}\fb}, the Fermi energy {\bb \mmu\fb} is
determined as an integral over the first B-Z by
{\be \nn_{\rmn} = \frac{2}{(2\pi)^3}\sum_{\alp}\int_{_{\rm BZ}} {\rm d}^3 \kk\, \vartheta\{\mmu
- {\calE}_\alp \} \, ,\fe} where we have introduced the Heaviside unit step
distribution defined by
{\bb \vartheta\{ x\}=1\fb} if {\bb x>0\fb} and zero otherwise (the factor of 2 is
to account for the spin degeneracy). The conduction neutron density is defined by
the occupied single particle states whose energy is positive, i.e. {\bb {\calE}_{\alp}>0\fb}, namely
{\be \nn = \frac{2}{(2\pi)^3}\sum_{\alp}\int_{_{\rm BZ}} {\rm d}^3 \kk\,
\vartheta\{\mmu -{\calE}_\alp\}\vartheta\{ {\calE}_\alp \} \, .\fe}

It will be instructive to compare the Fermi surface area with that of a non
interacting neutron gas of density {\bb \nn_{\rmn}\fb}, which is given by
{\be {\sS }_{\rm gas}=4\pi(3\pi^2\nn_{\rmn})^{2/3}\, .\fe}

\subsection{``Lasagna'' phase}
In this model, the slab shaped nuclei or ``lasagna'' are parallel to each
other and equally spaced by a distance {\bb a\fb}. Hence the
single particle potential {\bb \VV\fb} is periodic along one direction only,
say the {\bb z\fb} axis, and takes a constant value in the other two
dimensions. The neutron band structure with a square well potential
(whose analytic solution was found a long time ago by Kronig and
Penney \cite{KronigPenney,Folland}) was the only case discussed by
Oyamatsu and Yamada \cite{Oyamatsu94}.

The single particle wavefunction can be factored in the form
{\be \phhi_{\bf \kk }\{ {\bf r } \} = \phi_{\kk_z}\{ z\}{\rm e}^{{\rm i}\,
(\kk_x x + \kk_y y)}\, , \fe}
in which the reduced wave funtion {\bb \phi_{ \kk_z}\fb} is the solution of
a one dimensional equation of the form
{\be \frac{-\hba^2}{2 \mm}\frac{d^2 \phi_{\kk_z}}{dz^2} + \VV\{ z\}
\phi_{\kk_z}\{ z\} = \varepsi\{ \kk_z\} \phi_{\kk_z}\{ z \}\, , \fe}
satisfying the boundary conditions
{\be \phi_{\kk_z}\{ z+ a \} = e^{{\rm i}\,  \kk_z a } \phi_{\kk_z}\{ z \}
\, .\fe}

The first B-Z is defined by the set of Bloch wave vectors such
that {\bb -\pi/a\leq \kk_z\leq \pi/a.\fb} The
B-Z in this peculiar case has an infinite extent as a result of the continuous
translational invariance along planes parallel to the ``lasagna''.
The single particle energy is thus given by an expression of the form
{\be {\calE}_{\alp}\{ {\bf \kk } \} = \frac{\hba^2 (\kk_x^2 + \kk_y^2)}
{2 \mm}+\varepsi_{\alp}\{ \kk_z\},\fe} where {\bb \alp \fb} is a band index.
It follows that the neutron group velocity components parallel to the slabs
coincide with the group velocity of a non interacting neutron gas while only the
component perpendicular to the slabs is affected. It can be shown
from group theoretical arguments, that this component of the group velocity
 must vanish at the Brillouin zone edge \cite{Altmann}, i.e. at {\bb \kk_z=
\pm\pi /a\fb}. More generally whenever the Bloch wave vector lies on a symmetry
plane, the group velocity component normal to the plane vanishes .
It can also be shown that the energy bands do not cross \cite{Altmann}(nevertheless bands
may touch ``accidentally'' for some specific
choice of potential, for instance a constant one).

The total neutron density is given by
{\be \nn_{\rmn} = \frac{\mm}{\pi^2 \hba^2}\sum_{\alp} \int_0^{\pi/a} {\rm d}
\kk_z \bigl(\mmu-\varepsi_{\alp}\bigr) \vartheta\{ \mmu-
\varepsi_{\alp} \}\, .\fe}

The mobility along the ``lasagna'' planes coincides with that of the
uniform neutron gas with the same total neutron density:
 {\be {\calK}^{\parallel}={\calK}^{xx} = {\calK}^{yy} = \frac{\nn_{\rmn}}{\mm}\, .\fe}

The mobility in directions perpendicular to the ``lasagna'' may however
differ from that of the non interacting neutron gas:
 {\be {\calK}^{\perp}={\calK}^{zz} = \frac{1}{4 \pi^3 \hba} \sum_{\alp}
\int \frac{(\vv^z)^2}{\vv} {\rm d}\sS_{_{\rm F}}^{(\alp)}=
\frac{\mm}{\pi^2 \hba^4}\sum_{\alp}\int_0^{\pi/a} \biggl(\frac
{d\varepsi_{\alp}}{d\kk_z}\biggr)^2 \vartheta\{ \mmu-
\varepsi_{\alp}\} {\rm d}\kk_z\, .\fe}

Since the ``lasagna'' configuration is strongly anisotropic, it is more
appropriate to define a transverse effective macro mass by
{\be \mm_{\star}^{\perp} \equiv \frac{\nn}{{\calK}^{\perp}}. \fe}

The conduction neutron density is expressible as
{\be \nn = \nn_{\rmn}+\frac{4}{(2\pi)^3}\sum_{\alp}\int_{0}^{\pi/a}
\pi \frac{2 \mm \varepsi_\alp}{\hba^2} \vartheta\{ \mmu-
\varepsi_{\alp}\} \vartheta\{ -\varepsi_\alp\} {\rm d}\kk_z \, .\fe}

\subsection{``Spaghetti'' phase}
In the spaghetti model, the crystal is composed of cylinder shaped nuclei
arranged on a two dimensional lattice (the nuclear ``spaghetti'' are assumed
to be parallel to each other along say the z axis).
Since the single particle potential around one isolated nucleus only depends
on the distance to the rod like nucleus, the corresponding contribution to
the crystal potential does not depend on {\bb z\fb} hence the wave function
can be factored as
{\be \phhi_{\bf \kk }\{{\bf r }\} = \phi_{\kk_x, \kk_y}\{x,y\}
{\rm e}^{{\rm i}\, \kk_z z} \, .\fe}
The two dimensional wave function
obeys a Schr\"odinger equation of the form
{\be -\frac{\hba^2}{2 \mm}\biggl(\frac{\partial^2}{\partial x^2}+
\frac{\partial^2}{\partial y^2}\biggr) \phi_{\kk_x,\kk_y} \{x,y\} +
\VV\{x,y\}\phi_{\kk_x, \kk_y}\{x,y\} = \varepsi\{ \kk_x,\kk_y\}
\phi_{\kk_x,\kk_y}\{x,y\}\, . \fe}
The energy is thus decomposible as
{\be {\calE}_{\alp}\{ {\bf \kk } \} =\varepsi_{\alp}\{ \kk_x,\kk_y \}
+\frac{\hba^2 \kk_z^2}{2 \mm}\, .\fe}

The total neutron density is given by
{\be \nn_{\rmn} = \frac{4}{(2\pi)^3 \hba}\sum_{\alp} \int_{_{\rm BZ}}
\sqrt{ 2 \mm (\mmu-\varepsi_{\alp})} \, \vartheta\{\mmu
-\varepsi_{\alp}\}{\rm d}\kk_x {\rm d}\kk_y \, ,\fe}
in which the factor of two arises from the
restriction that {\bb \kk_z>0\fb} and where the integration is carried out
over the 2D first B-Z illustrated on figure
\ref{fig7}.

It is readily verified that the mobility tensor component along the
cylindrical nuclei is merely equal to the mobility component of the non interacting
neutron gas:
{\be {\calK}^{\parallel} ={\calK}^{zz} = \frac{\nn_{\rmn}}{\mm}\, .\fe}
 The other components perpendicular to the ``spaghetti'' depend on the
neutron-crystal interaction. It is convenient to defined a mean transverse
mobility by
{\be {\calK}^\perp =  \sqrt{\frac{\mm}{2}}\frac{2}{(2 \pi)^3 \hba^3}
\sum_{\alp}\int_{_{\rm BZ}} \frac{ {\rm d}\kk_x {\rm d}\kk_y}{\sqrt{\mmu
-\varepsi_{\alp}}} \left( \biggl(\frac{\partial \varepsi_{\alp}}
{\partial \kk_x} \biggr)^2 +\biggl(\frac{\partial \varepsi_{\alp}}
{\partial \kk_y} \biggr)^2 \right)  \, \vartheta\{ \mmu-\varepsi_{\alp}\}, \fe}
since it involves the integral of a completely symmetric function that can thus be
factored into one irreducible domain of the first B-Z \cite{Johnston60}.

\begin{figure}
\centering
\epsfig{figure=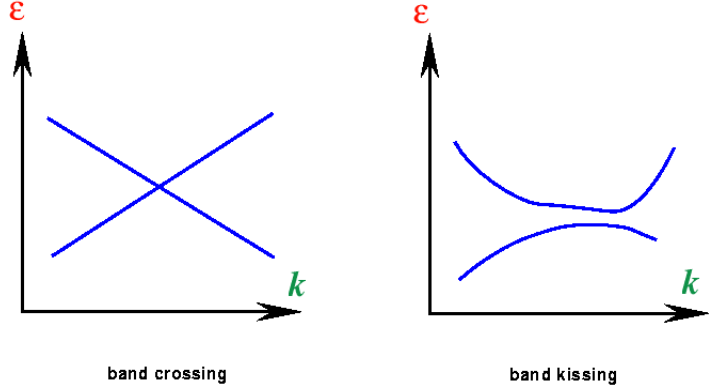, width=10 cm}
\caption{Sketch of energy against wave number for crossing and ``kissing''
scenarios.
\label{fig6}}
\end{figure}

\begin{figure}
\centering
\epsfig{figure=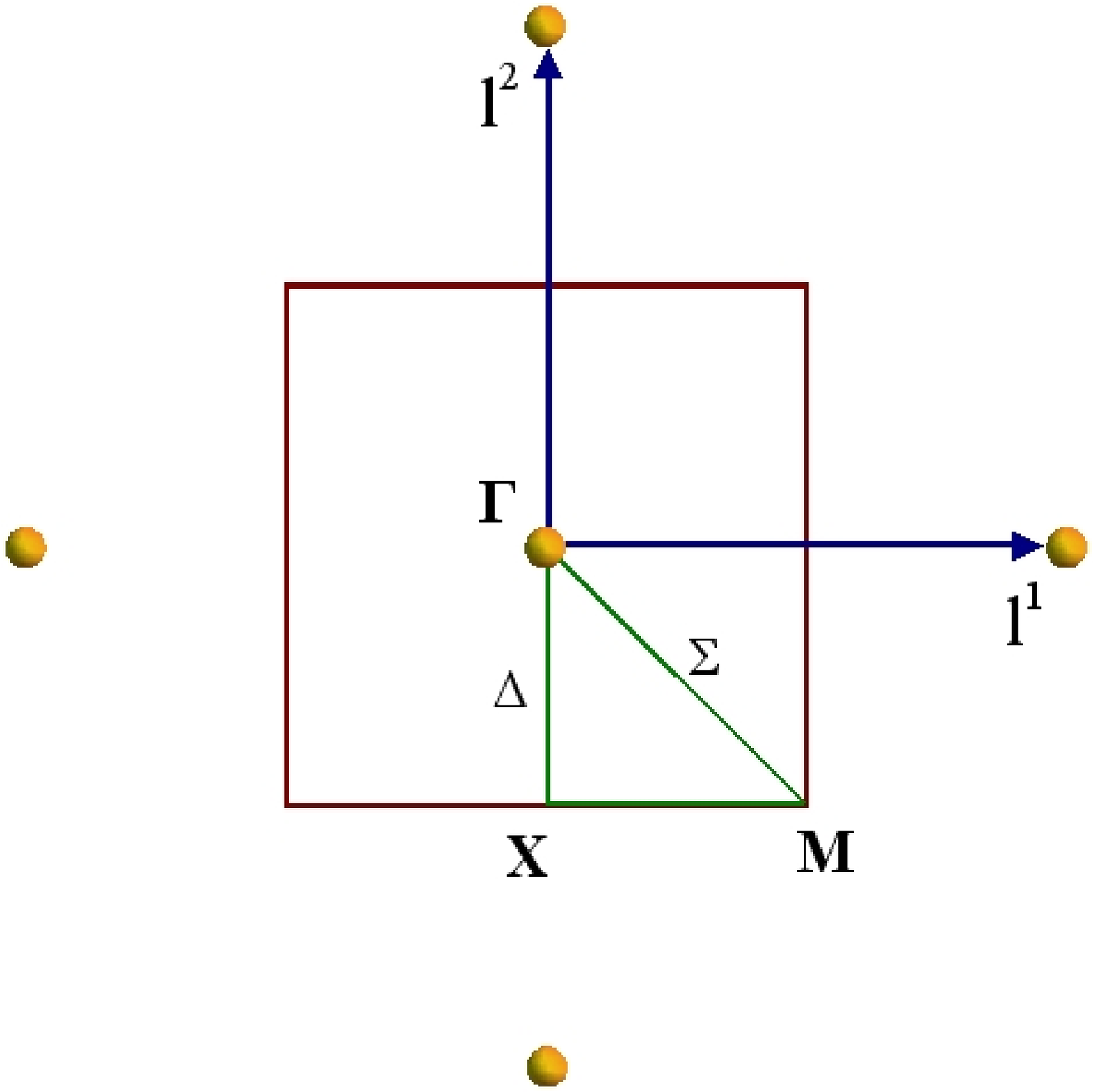, width=5 cm}\hskip 1cm\epsfig{figure = 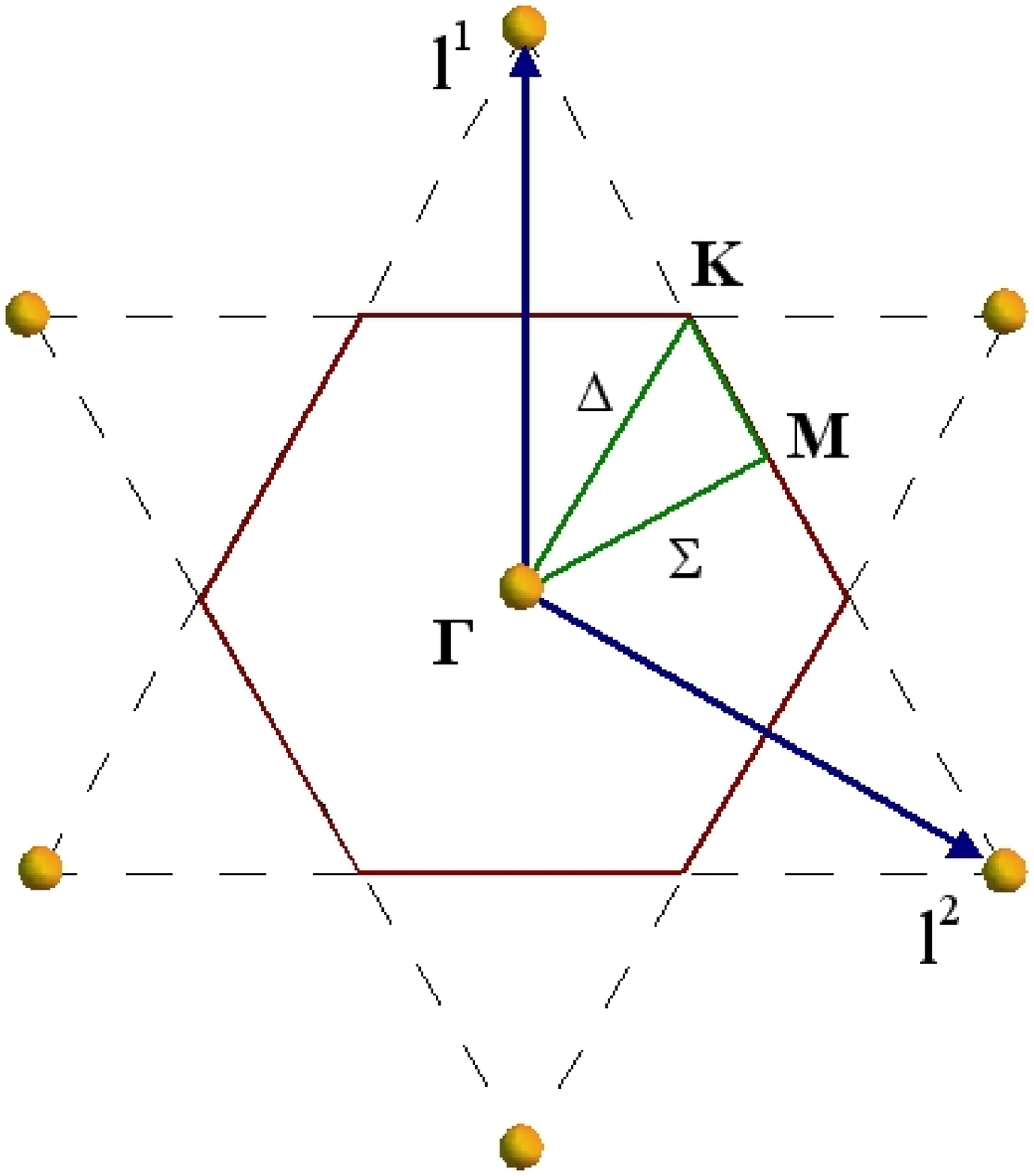, width=5 cm}
\caption{First Brillouin zones and irreducible domains for square and
hexagonal variants of spaghetti (rod) lattice.
\label{fig7}}
\end{figure}

The Fermi surface area, the mobility tensor and the
density involve integrations over the two dimensional first B-Z of
functions weighted by a Fermi distribution. We have followed an idea due to
Gilat and Raubenheimer for three dimensional crystals \cite{GilatRaubenheimer},
and translated it into the two dimensional case. First of all the zone is
decomposed into small identical cells within which the single particle energy
is linearly extrapolated from the value of the energy and its gradient at the
center of the cell. The integration is then performed analytically inside the
cell. The integral is approximated by summing the contribution from each cell
(properly weighted whenever the cells overfill the zone). Unlike the case of
band crossing, extrapolation fails for band ``kissing'' where the energy
gradient varies rapidly in the cell (see the schematic pictures on Figure
\ref{fig6}). This problem becomes more and more acute as the number of bands
to be included in the integration is increased. However those induced
systematic errors will tend to vanish as the number of microcells is
increased. Whereas the integrand could be integrated analytically inside each
microcell, we found that it is better to take it as constant. The reason
again lies in the band ``kissing'' problem. The integrand typically depends on
the momentum via the single particle energy which is extrapolated. Near a
band kissing region, the extrapolation errors in the integrand are integrated
which lead to unstable results.

We have considered two lattice types: square and hexagonal crystals (whose
reciprocal lattices are also square and hexagonal respectively). The point
group {\bb \cal P\fb} (the set of point symmetries which send the lattice
into itself) of the hexagonal lattice is {\bb C_{6v}\fb}
(Sch\"onflies notation \cite{Hamermesh}) whose order is equal to {\bb |{\cal P}|=12,\fb}
 thereby reducing integrations of completely symmetric functions over the
entire two dimensional first B-Z to {\bb 1/12^{\rm th}\fb} of the
zone. We mention that for this structure the lattice
spacing {\bb a\fb} (shortest distance between any two lattice points) is not
equal to the value given by Oyamatsu {\bb a_{_{\rm Oy}}\fb} but is given by
{\be a=\sqrt{\frac{2}{\sqrt{3}}}\, a_{_{\rm Oy}}\, . \fe}
Likewise, the first B-Z of a square lattice whose point group is
{\bb C_{4v}\fb} (lattice spacing {\bb a=a_{_{\rm Oy}}\fb}) can be
partitionned into {\bb 8\fb} irreducible domains. These considerations are illustrated in
Figure \ref{fig7} with the conventional labelling (first B-Z in
red, one irreducible domain in green). The most adequate cell for
integrations is thus a rectangle since the irreducible B-Zs are
rectangular triangles. The partition of this domain into microcells is
thereby straightforward.

\begin{figure}
\centering
\epsfig{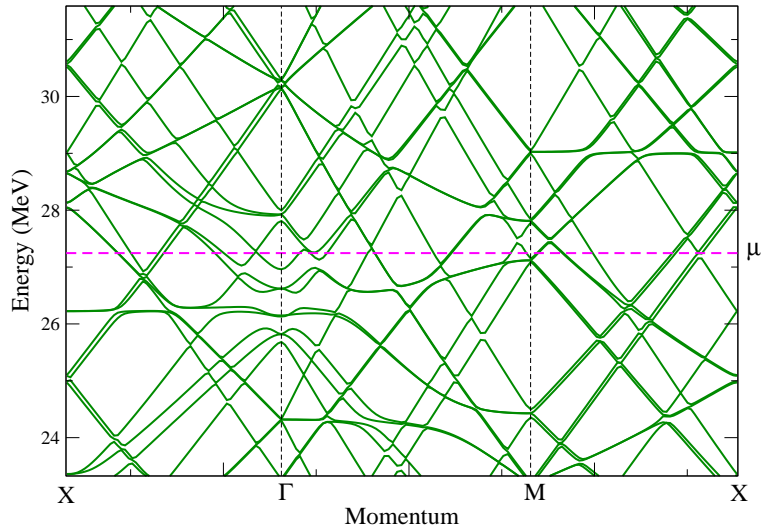}
\caption{Neutron band structure around the Fermi energy {\bb \mmu\fb},
along directions shown on figure \ref{fig7} for square variant of spaghetti
(rod) lattice ({\bb a_{_{\rm Oy}}=27.17\fb} fm).
\label{fig8}}
\end{figure}

\begin{figure}
\centering
\epsfig{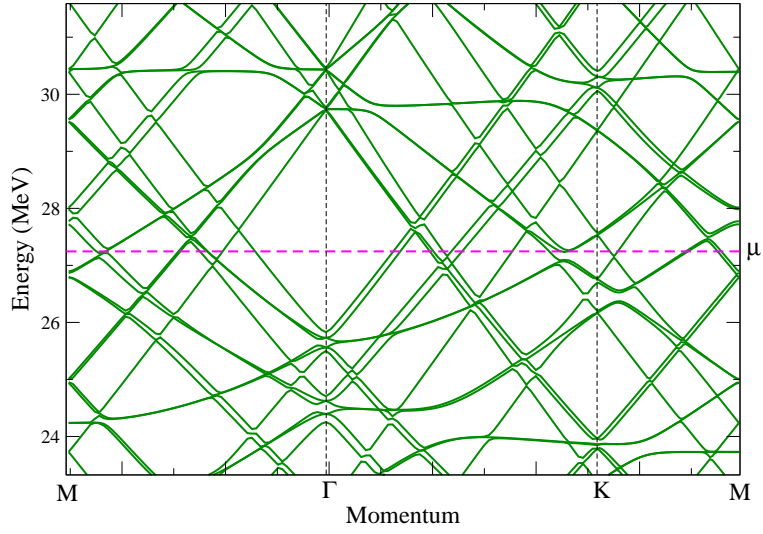}
\caption{Neutron band structure around the Fermi energy {\bb \mmu\fb},
along directions shown on figure \ref{fig7} for hexagonal variant of spaghetti
(rod) lattice with {\bb a_{_{\rm Oy}}=27.17\, \rm fm\, .\fb}
\label{fig9}}
\end{figure}

These two groups, {\bb C_{6v}\fb} and {\bb C_{4v}\fb}, contains two
dimensional irreducible representations \cite{Hamermesh} which means that
unlike the 1D case, energy bands may cross each other \cite{Altmann}. The neutron band
structure along high symmetry lines is shown on Figures \ref{fig8} and
\ref{fig9} for energies in the vicinity of the Fermi energy and for the
lattice spacing {\bb a_{_{\rm Oy}}=27.17\, \rm fm\, .\fb}

The conduction neutron density is equal to
{\be \nn = \nn_{\rmn}-\frac{4}{(2\pi)^3 \hba}\sum_{\alp} \int_{\rm BZ}
\sqrt{ -2 \mm \varepsi_{\alp}} \, \vartheta\{\mmu
-\varepsi_{\alp}\}\vartheta\{-\varepsi_\alp\}{\rm d}\kk_x {\rm d}\kk_y \, .\fe}
The transverse effective mass is defined by
{\be \mm_{\star}^{\perp} \equiv \frac{\nn}{{\calK}^{\perp}}.\fe}

\begin{figure}
\centering
\epsfig{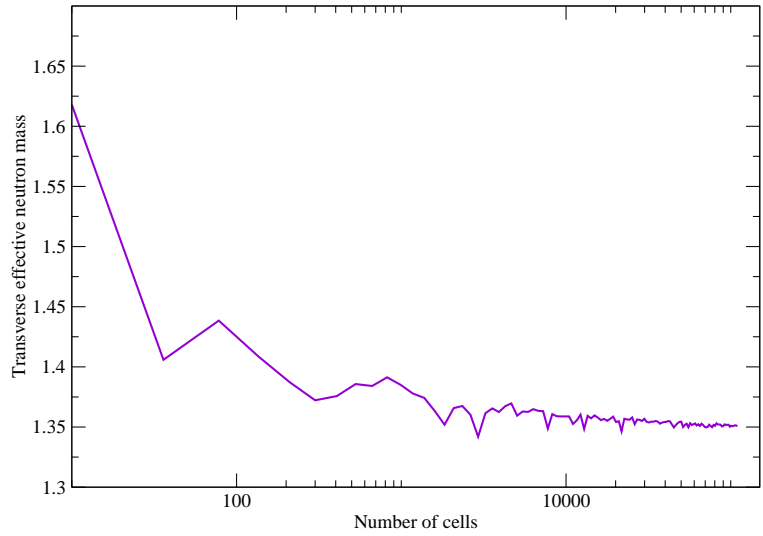}
\caption{Convergence of computed effective mass {\bb \mm_\star^{\perp}/\mm\fb} as a function of cell number for an hexagonal lattice
with {\bb a_{_{\rm Oy}}=27.17\, \rm fm\, .\fb}
\label{fig12}}
\end{figure}

\begin{figure}
\centering
\epsfig{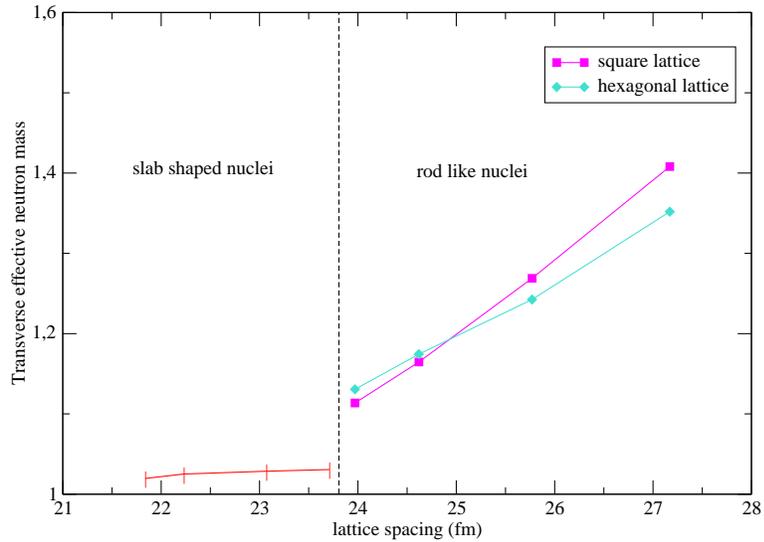}
\caption{Effective mass of conduction neutrons {\bb \mm_\star^{\perp}/\mm\fb} as a function of lattice
spacing ({\bb a_{_{\rm Oy}}\fb}) below and above phase transition from lasagna
(slab) regime to spaghetti (rod) regime.
\label{fig10}}
\end{figure}

 The convergence of the integration scheme based on a decomposition into
rectangular cells is illustrated on figure \ref{fig12} for the hexagonal
lattice with the lattice spacing {\bb a_{_{\rm Oy}}=27.17\, \rm fm\, .\fb} The convergence
 is much faster for the density than for the other quantities due to the
absence of the singular square root integrand.

\section{Discussion}

\begin{figure}
\centering
\epsfig{figure=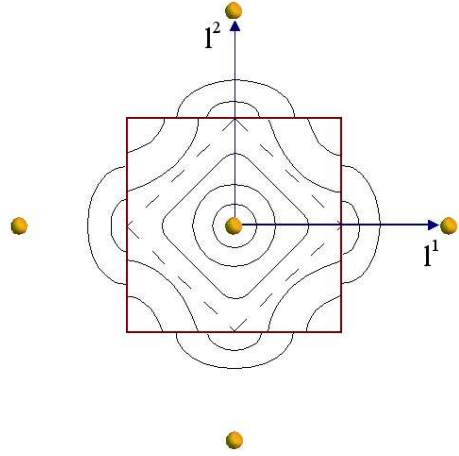, width=6 cm}
\caption{Example of energy contours in extended Brillouin zone for square
lattice. For sufficiently small Fermi energies the contours are
approximately circular as in the free particle case, while at higher energies, the contours
consist of disconnected pieces.
\label{fig2}}
\end{figure}

The results concerning the macroscopic effective neutron mass are illustrated on figure \ref{fig10} and numerical
values can be found in the appendix. The macroscopic effective neutron mass appears to be increased compared to the
ordinary neutron mass. The figure also shows that this renormalisation of the neutron mass is mostly significant at low densities and becomes negligible at higher densities where nuclei nearly merge
into a uniform mixture. The deviations of the effective neutron mass from the bare one can be understood in terms of modifications of the Fermi surface from a sphere. In particular,
as a result of the opening of band gaps the Fermi surface is not smooth but contains many holes as schematically illustrated on figure \ref{fig2}. Since the enclosed Fermi volume only depends
on the density, this means that the Fermi surface area for a given density is reduced as compared to the corresponding sphere as shown on figure \ref{fig11} (see the appendix for the numerical values).
In the case for which the Fermi volume is equal to the volume of the first Brillouin zone, the Fermi sphere can be distorted and torn such that the resulting surface area simply vanishes while the
volume remains finite. Such a situation occurs in ordinary electric insulators.

 \begin{figure}
\centering
\epsfig{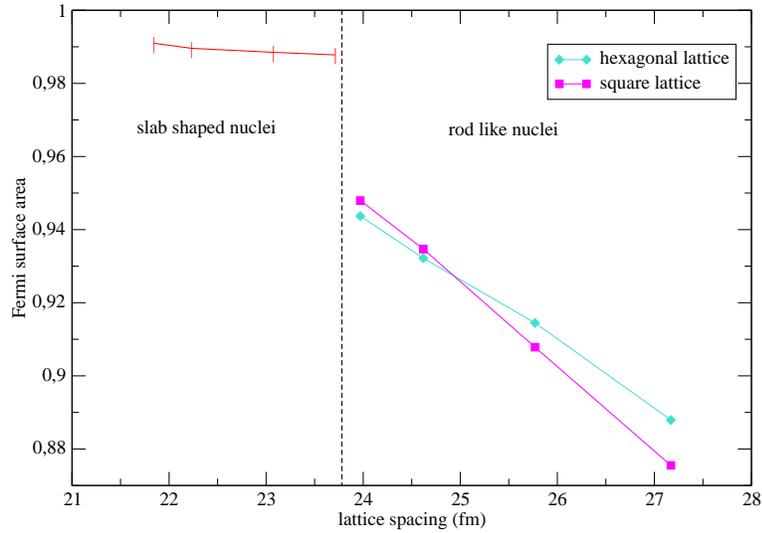}
\caption{Fermi surface area for neutrons {\bb {\sS }_{_{\rm F}}/
{\sS }_{\rm gas}\fb} as a function of lattice spacing ({\bb a_{_{\rm Oy}}\fb})
below and above phase transition from lasagna (slab) regime to spaghetti
(rod) regime.
\label{fig11}}
\end{figure}

In the present work we have neglected the spin-orbit coupling which is a small perturbation
in the density region we considered. Taking into account such
term would raise some degeneracies. The most dramatic change
 concerns some regions in momentum space around which unperturbed bands are
crossing. The breaking of the spin symmetry will entail that such
configuration may be turned into band ``kissing'' (see figure \ref{fig6})
which means that the group velocity will be strongly reduced around those momenta. This
 means that the resulting mobility would be lower than the one we have found.
 Besides, since the conduction neutron density is rather unsensitive to such change,
 it would not be much affected and therefore the effective neutron mass would be
 slightly larger than the one calculated here.

\section{Conclusion}

The scattering of dripped neutrons by the nuclei in the inner crust leads on
a macroscopic scale
to a modification of the neutron mass
{\bb  \mm_ \star \fb}, which can be expressed
 via a well defined mobility scalar {\bb  \calK \fb} by {\bb  \mm_\star
=\nn/\calK \fb} in which {\bb  \nn \fb} is the (arbitrary) density of such unbound neutrons,
and {\bb  \calK \fb} is
found to be expressible as an integral of the group velocity {\bb  \vv^i \fb} over
the corresponding Fermi surface. This effective macro mass should not be confused
with the effective micro mass, relevant for subnuclear scales,
 which is usually found to be smaller than the ordinary neutron mass.

Bragg scattering of dripped neutrons is taken into account here by applying Bloch type
boundary conditions, which are well known in solid state physics but have been barely used
in this nuclear context. We have computed numerical values of this mobility scalar in the bottom
layers of the inner crust near the crust-core interface, for simple models
 in the ``pasta'' layers: equally spaced slab shaped nuclei
(``lasagna'') and rod like nuclei on either a square or an hexagonal 2D lattice
(``spaghetti''). The (anisotropic) entrainment effect is small at such
densities since the system is nearly homogeneous. It appears that the mobility
scalar tends to be systematically reduced compared to the homogeneous
expression and the farther from homogeneity, the smaller is the mobility
scalar. The resulting effective mass {\bb  \mm_\star \fb} is found to be larger
than the bare neutron mass. This results can be interpreted as a macroscopic manifestation of the
modifications in the shape of the Fermi surface. Notably as one goes from the homogeneous
outer core to the crust, the spherical neutron Fermi surface gets distorted and even torn and pierced.

The main reason is that the neutron Bragg scattering by crustal nuclei leads to the opening of
energy band gaps. A specific feature
of those exotic phases is that the mobility scalar is bounded,
{\bb \calK \geq 2\nn_{\rmn} / 3 \mm \fb} for the ``lasagna'' phase
and {\bb \calK \geq \nn_{\rmn} / 3 \mm \fb} for the ``spaghetti'' phase
due to the fact that the neutrons are still free to move in one or two
dimensions, respectively. There is no such bound for three dimensional
crystals, for which  smaller mobility scalars (larger effective masses) may be
expected.

From these considerations, we can infer that the mobility scalar {\bb  \calK \fb} will 
increase with the density, starting from zero in the outer crust below the
neutron drip threshold where all neutrons are confined, since {\bb  \vv^i=0 \fb} for all 
states,  to its largest possible value in the homogeneous neutron star mantle. At low 
densities near neutron drip, where the conduction neutron density is negligible compared to 
the total neutron density, {\bb \nn \ll \nn_{\rmn}\, ,\fb}  the crystal potential {\bb  \VV\fb} 
is quite large but the fraction of the cell occupied by the nuclei is small so that
dripped neutrons propagate essentially freely. Consequently we may expect to get
{\bb  \calK \simeq \nn/\mm\fb} and {\bb  \mm_\star\simeq \mm\fb} 
(on the understanding that, as discussed in Subsection \ref{cutoff}, a conduction state is 
defined so as to have an associated group velocity that differs significantly from zero).
 On the other hand,  at very high densities where nuclei nearly merge, the potential is 
much weaker and smoothly varying and we shall have {\bb \nn\simeq \nn_{\rmn}\, ,\fb}
 so we expect to find {\bb  \mm_\star\simeq \mm\, .\fb}

In other words the mean effective mass {\bb  \mm_\star\fb} is expected to be
close to the ordinary mass {\bb  \mm\fb} at the top (above neutron drip density where 
neutrons start to leak out of nuclei)
and bottom (where nuclei merge into a uniform mixture of neutrons, protons and
electrons at density about $1/3$ or $2/3$ the nuclear saturation density
{\bb  \nn_{\rm sat}\simeq 0.16\, {\rm fm}^{-3}\fb} \cite{Haensel01})
of the inner crust, but may reach a much larger value (with astrophysically interesting consequences)
in the intermediate layers. This indicates
that the unbound neutron band effects
(that are effectively neglected in the  commonly
used W-S approximation) could have important consequences
as concerns the equilibrium neutron star crust structure and composition.

The exploratory character of the present work, justifies the simplicity of the single particle model we used. In a more realistic, Hartree-Fock calculation,
the single particle potential and effective micro mass should have to be determined self-consistently. The potential may contain momentum dependent terms,
as with effective nucleon-nucleon interactions of the Skyrme type, which leads to space varying effective neutron masses {\bb \mm^{_\oplus}\{ {\bf r} \} \fb}
which are typically smaller than the bare neutron mass. However the general arguments we developed, relying on the shape of the Fermi surface,
suggest that the qualitative results of our calculations, namely the enhancement of the macroscopic effective mass {\bb \mm_\star \fb} will remain
valid in more elaborate single particle schemes. Finally let us mention that as we have shown in a recent paper \cite{CCHIII}, the neutron pairing is not expected
to qualitatively alter our present conclusions.

\vskip 0.5cm
{\bf Acknowledgements}

One of the authors (PH) was partly supported by the
KBN grant no. 1-P03D-008-27
and by the LEA Astro-PF PAN/CNRS program.
We wish to thank the referee for questions that have helped us to improve the clarity of our presentation.

\vfill\eject
{\bf APPENDIX}

\appendix
\section{Numerical results}
\subsection{Slab shaped nuclei}
\begin{tabular}{|p{1.9cm} || p{1.7cm} | p{1.5cm} | p{1.7cm} | p{1.7cm} |
 p{1.7cm}| p{1.7cm} |  }
\hline
 {\bb \nn_{\rmn}\, ({\rm fm}^{-3})\fb}  & {\bb \mmu({\rm MeV})\fb} & {\bb a_{\rm Oy}
({\rm fm})\fb} & {\bb \nn/\nn_{\rmn}\fb} & {\bb {\calK}^{^{\parallel}}/
{\calK}^{^{\perp}}\fb} & {\bb \mm_{\star}^{\perp}/\mm\fb}
& {\bb {\sS }_{_{\rm F}}/{\sS }_{\rm gas}\fb}
  \\ \hline
 0.0735 & 31.70 & 23.71 &  0.9634 & 1.0698& 1.0307 & 0.9878
 \\ \hline
 0.0749 & 32.10 & 23.07 & 0.9646 & 1.0664 & 1.0286 & 0.9885
  \\ \hline
 0.0773  & 32.79 & 22.23 & 0.9666 & 1.0605 & 1.0251 & 0.9896
 \\ \hline
 0.0792  & 33.36 & 21.84 & 0.9687 & 1.0526   & 1.0196 & 0.9910
 \\ \hline
\end{tabular}

\subsection{Rod like nuclei}

\subsubsection{hexagonal lattice}
\begin{tabular}{|p{1.9cm} || p{1.7cm} | p{1.5cm} | p{1.7cm} | p{1.7cm}| p{1.7cm}| p{1.7cm}| }\hline
  {\bb \nn_{\rmn}\, ({\rm fm}^{-3})\fb}  & {\bb \mmu ({\rm MeV})\fb} & {\bb a_{\rm Oy} ({\rm fm})\fb}& {\bb \nn/\nn_{\rmn}\fb} & {\bb {\calK}^{\parallel}/{\calK}^{\perp}\fb}
& {\bb \mm_{\star}^{\perp}/\mm\fb} & {\bb {\cal S}_{_{\rm F}}/{\cal S}_{\rm gas}\fb}    \\ \hline
 0.0581 & 27.2461 & 27.17 & 0.9553 & 1.4102 & 1.3471 & 0.8898    \\ \hline
0.0630 & 28.7422 & 25.77 & 0.9578 & 1.2972 & 1.2425 & 0.9145    \\ \hline
0.0678 & 30.1836 & 24.62 & 0.9606 & 1.2225 & 1.1744 &  0.9322    \\ \hline
0.0716 & 31.2812 & 23.97 & 0.9632 &1.1668  & 1.1239 & 0.9471   \\ \hline
\end{tabular}

\subsubsection{square lattice}
\begin{tabular}{|p{1.9cm} || p{1.7cm} | p{1.5cm} | p{1.7cm} | p{1.7cm}| p{1.7cm}| p{1.7cm}| p{1.7cm}| }\hline
  {\bb \nn_{\rmn}\, ({\rm fm}^{-3})\fb}  & {\bb \mmu ({\rm MeV})\fb} & {\bb a_{\rm Oy} ({\rm fm})\fb}& {\bb \nn/\nn_{\rmn}\fb} & {\bb {\calK}^{\parallel}/{\calK}^{\perp}\fb}
& {\bb \mm_{\star}^{\perp}/\mm\fb} & {\bb {\cal S}_{_{\rm F}}/{\cal S}_{\rm gas}\fb}   \\ \hline
0.0581 & 27.2461 & 27.17 & 0.9553 & 1.4673  & 1.4016 & 0.8778  \\ \hline
0.0630 & 28.7422 & 25.77 & 0.9578 & 1.3231 & 1.2673 & 0.9085    \\ \hline
0.0678 & 30.1797 & 24.62  & 0.9606 & 1.2124 & 1.1647  & 0.9347   \\ \hline
0.0716 & 31.2812 & 23.97 & 0.9632 & 1.1473  & 1.1051 & 0.9524    \\ \hline
\end{tabular}

\end{document}